\documentclass[reprint,superscriptaddress,floatfix,aps,prl]{revtex4-2}

\usepackage{header}

\begin{document}
\title{Detecting nonequilibrium phase transitions via continuous monitoring of space-time trajectories and autoencoder-based clustering}
\author{Erik Fitzner}
\affiliation{Institut f\"ur Theoretische Physik, Universit\"at T\"ubingen, Auf der Morgenstelle 14, 72076 T\"ubingen, Germany}
\author{Francesco Carnazza}
\affiliation{Universit\'{e} Paris Cit\'e, CNRS, Mat\'{e}riaux et Ph\'{e}nom\`{e}nes Quantiques, 75013 Paris, France}
\author{Federico Carollo}
\affiliation{Centre for Fluid and Complex Systems, Coventry University, Coventry CV1 2TT, United Kingdom}
\author{Igor Lesanovsky}
\affiliation{Institut f\"ur Theoretische Physik, Universit\"at T\"ubingen, Auf der Morgenstelle 14, 72076 T\"ubingen, Germany}
\affiliation{School of Physics and Astronomy and Centre for the Mathematics and Theoretical Physics of Quantum nonequilibrium Systems, The University of Nottingham, Nottingham, NG7 2RD, United Kingdom}

\begin{abstract}
    The characterization of collective behavior and nonequilibrium phase transitions in quantum systems is typically rooted in the analysis of suitable system observables, so-called order parameters. These observables might not be known a priori, but they may in principle be identified through analyzing the quantum state of the system. Experimentally, this can be particularly demanding as estimating quantum states and expectation values of quantum observables requires a large number of projective measurements. However, open quantum systems can be probed {\it in situ} by monitoring their output, e.g.~via heterodyne-detection or photon-counting experiments, which provide space-time resolved information about their dynamics. Building on this, we present a machine-learning approach to detect nonequilibrium phase transitions from the measurement time-records of continuously-monitored quantum systems. We benchmark our method using the {\it quantum contact process}, a model featuring an absorbing-state phase transition, which constitutes a particularly challenging test case for the quantum simulation of nonequilibrium processes. 
\end{abstract}

\maketitle

% -------------------------------------------------------------------------------------------------------------------------------------------- %

\textit{Introduction.---} Interacting many-body quantum systems far from equilibrium can display complex collective behavior and phase transitions~\cite{Non_Eq,Rossini2021,Landi2022,Skinner2019,Igor2013,Beaulieu2025,Shah2025,Walter2014,carollo2022,Sacha2018,Wilczek2012,Nahum2017}. The analysis of these phenomena is typically based on the identification of a suitable observable, a so-called {\it order parameter}, which captures the abrupt changes to the system state occurring close to a critical point. Their characterization thus effectively relies on a {\it dimensional reduction}, whereby collective effects and universal properties are fully described by the behavior of a single quantity. From a theoretical viewpoint, order parameters can be obtained by calculating expectation values of system observables or even by considering more involved properties of the quantum state, such as entanglement and quantum correlations. However, determining these quantities experimentally may come with significant overhead, which is associated with repeated state preparation and projective measurements or, in the most extreme case, with a full quantum state tomography.
 
\begin{figure}
    \centering
    \includegraphics[width=0.95\linewidth]{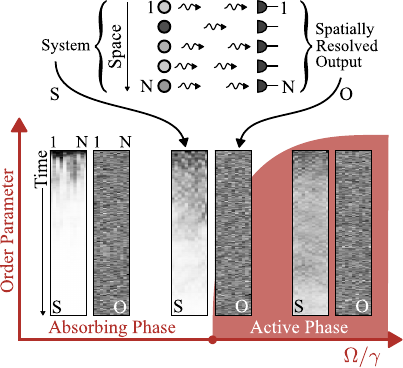}
    \caption{\textbf{Quantum trajectories across the phase diagram of the quantum contact process.} The top part of the figure sketches an experimental setting in which spatially resolved emissions from a many-body system (S) are monitored continuously in time. The sketched quantum trajectories labeled by S show qualitatively the dynamics of the expectation of a local order parameter (here the local density of active sites). The latter display clearly distinct structures in the subcritical, critical, and supercritical regimes of the quantum contact process. In contrast, space-time records O of the output from continuous monitoring [here the real part of the complex heterodyne current, Eq.~(\ref{eq:Jhet})] appear noisy and structureless throughout all dynamical regimes. While trajectories of the order parameter are usually inaccessible due to postselection overheads, trajectories of the output signal are directly available in experiments. As we show, they can provide useful information to detect nonequilibrium phase transitions.}
    \label{fig:sing_traj}
\end{figure}

Continuously monitored open quantum systems release information into the environment which can be in situ monitored and processed. This space-time resolved information (see sketch in Fig.~\ref{fig:sing_traj}) can be obtained using heterodyne or photon‐counting measurements~\cite{breuer2002theory,gardiner2004quantum,QM_Measurement,jin2025}, using digital quantum simulators~\cite{Guo2024,Impertro2024,Bluvstein2024,Singh2023,Koh2023,Fauseweh2024,Bernien2017,Bloch2012}, or via ancilla‐based measurement schemes~\cite{Marcel_ancilla,anand2024}. In contrast to projective measurements, continuous monitoring does not destroy the state of the system. It rather  provides continuous dynamical information~\cite{QM_Measurement,Gisin}, encoded in so-called quantum trajectories (cf.~Fig.~\ref{fig:sing_traj}), which is experimentally accessible without the need of repeated state preparation, ensemble averaging, or postselection. 
Moreover, one may expect that the space-time information contained in quantum trajectories allows to distinguish dynamical phases and to pinpoint phase transitions.

\begin{figure*}[t]
    \centering
    \includegraphics[width=\linewidth]{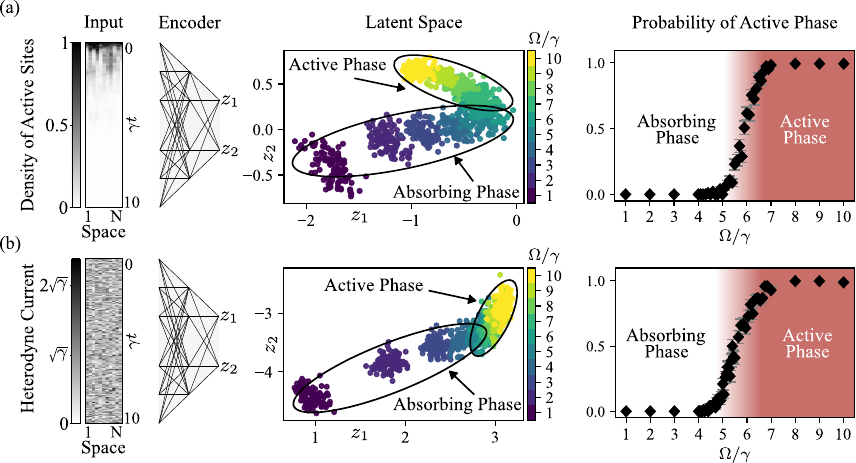}
    \caption{\textbf{Classification of quantum trajectories.} Individual trajectories of the one-dimensional quantum contact process with $N=30$ sites are classified using an autoencoder. The trajectories are encoded by two latent variables $z_1$, $z_2$, which span the latent space where trajectories from the active and absorbing phases separate. As mentioned in the main text, the latent variables are assumed to follow a bimodal multivariate Gaussian (sketched in in the central panel), which allows to assign to each encoded trajectory a probability of belonging to the component associated with the active phase. (a) Latent space representation of $1000$ trajectories constructed from the local density of active sites $S_k(t)$. Note that such trajectories are not practically accessible in experiment albeit simple to compute in numerical calculations. The probability of belonging to the active phase exhibits a sharp crossover as a function of the control parameter $\Omega/\gamma$. (b) Corresponding analysis based on the absolute value of the time-averaged heterodyne current: $|\overline{O_k(t)}|$. These quantum trajectories can be in principle accessed in an experiment. For both analyses the density of data points in the interval $\Omega/\gamma \in [4.0,\,7.0]$ is increased to better resolve the transition region.}
    \label{fig:Clustering}
\end{figure*}

In this work, we introduce a machine-learning approach that learns collective properties of open quantum systems directly from the structure of quantum trajectories. 
While this output signal need not be directly related to any system observable or order parameter, we show that it encodes crucial information for the neural network to be able to discriminate between different phases. We achieve this by feeding the high‐dimensional space-time resolved quantum trajectories to an unsupervised learning framework, which operates by mapping them onto a low‐dimensional {\it latent space}. Here, trajectories are clustered according to the dynamical phase they originate from (see an example in Fig.~\ref{fig:Clustering}). 
In this way, we exploit the intrinsic capability of machine-learning methods to reduce the problem's dimensionality and to extract effective order parameters from quantum trajectories. This allows us to detect signatures of criticality from complex datasets, without relying on predefined system observables~\cite{bény2013,mehta2014,Wetzel2017,Mlsciences}. 

We illustrate our method focusing on the quantum contact process~\cite{buchhold,Marcuzzi2016,carollo2019,Kahng2021} and synthetically generate heterodyne-measurement quantum trajectories using tensor networks. We select this model because it features a nonequilibrium phase transition --- even in one dimension --- between an absorbing phase and an active phase. The presence of the absorbing phase renders studying the model extremely challenging and a benchmark problem for quantum hardware~\cite{carollo2019,chertkov2023}. 
We emphasize that machine-learning has been employed for the identification and classification of many-body phases in Bose-Hubbard quantum simulators~\cite{Bohrdt2019,Impertro2023,Bohrdt2023,Carrasquilla2017} and, at a theoretical level, also for the quantum contact process~\cite{Kahng2021}. However, in these works system observables with a direct connection to order parameters (density or density-density correlations) were considered, or measured. In contrast, our approach exploits the entire space-time record of the measured output, which is not linked in an apparent way with the order parameter.

%%%
%%%

\textit{Continuously monitored open quantum systems.---} 
We consider the setting illustrated in Fig.~\ref{fig:sing_traj}, in which the output of the electromagnetic field of an open quantum system is monitored. We deliberately specialize to heterodyne detection~\cite{QM_Measurement,Gisin}, since for the model considered below photon counting may be related to the order parameter. The stochastic evolution of the system state is given by the equation 
\begin{equation}\label{eq:SD}
    \!{\rm d}\!\ket{\psi}=
    \left[-iH_\mathrm{eff} {\rm d}t+\sum_{k=1}^N\left(L_k-\langle L_k\rangle\right){\rm d}\xi_k\right]\!\ket{\psi}\,, 
\end{equation}
which provides the increment of the quantum state in the infinitesimal time-interval ${\rm d}t$. Here, $N$ is the number of subsystems composing the many-body system, which are all independently monitored (see sketch in Fig.~\ref{fig:sing_traj}). The operators $L_k$ are the jump operators and encode how the measurement process affects the system dynamics. Here, we have $\langle L_k\rangle=\bra{\psi}L_k\ket{\psi}$ and the operator 
\begin{equation}
    H_\mathrm{eff}= H+i\sum_{k=1}^N\left(\langle L^\dagger_k\rangle L_k-\frac{1}{2}L^\dagger_k L_k-\frac{1}{2}\langle L^\dagger_k\rangle\langle L_k\rangle\right)
\end{equation}
represents an effective Hamiltonian accounting for the deterministic part of the dynamics. The noise terms are proportional to independent complex Wiener increments, $\{{\rm d}\xi_k\}$, obeying the Ito rules~\cite{Gisin}, 
\begin{equation}
    \begin{aligned}
        \mathbb{E}[{\rm d}\xi_k]=0\,, \quad  \mathbb{E}[{\rm d}\xi_k^2]=0\, ,\quad  \mathbb{E}[{\rm d}\xi_j^* {\rm d}\xi_k]=\delta_{jk}{\rm d}t\,.
        \end{aligned}
\end{equation}
Each realization of Eq.~\eqref{eq:SD} is associated with a specific space-time record of the measured complex heterodyne photocurrent (see Ref.~\cite{QM_Measurement} for details), which we denote as
\begin{equation}
O_k=\langle L_k\rangle+\frac{{\rm d}\xi_k}{{\rm d}t}\,.\label{eq:Jhet}
\end{equation}
Averaging over different dynamical realizations provides the ensemble-average state of the system $\rho=\mathbb{E}[\ket{\psi}\!\bra{\psi}]$, whose evolution is governed by a quantum master equation of Lindblad type~\cite{Lindblad1976,Kossakowski1976}, i.e., 
\begin{equation}\label{eq:LM}
   \dot{\rho} = -i[H,\rho] + \mathcal{D}[\rho] \,. 
\end{equation}
The first term on the right-hand side of Eq.~\eqref{eq:LM} solely contains the quantum-coherent contribution to the dynamics~\cite{QM_Griffiths}. The second term, explicitly given by 
\begin{equation}
    \mathcal{D}[\rho] = \sum_{k=1}^N \left( L_k \rho L_k^\dagger - \frac{1}{2} \{ L_k^\dagger L_k, \rho \} \right)\,,
\end{equation}
describes the average effects of the monitoring process.

Information about the dynamics of the expectation value of a system observable, $\langle A\rangle={\rm Tr}\left(\rho A\right)$, requires gathering statistics from several projective measurements. This means that for each considered time, one has to prepare the initial state several times, run the dynamics and perform measurements. On the contrary, a heterodyne trajectory, i.e.~the space-time resolved output of the continuous monitoring process [cf.~Eq.~\eqref{eq:Jhet}], is directly accessible in situ during a single experiment.

\textit{Quantum contact process.---} 
The model we consider consists of $N$ two-level systems which are associated with the orthonormal basis states $\{\ket{\bullet},\ket{\circ}\}$, representing sites being active and inactive, respectively. The jump operators are given by $L_k = \sqrt{\gamma}\, \sigma_k^-$, with $\sigma^-=\ket{\circ}\!\bra{\bullet}$ implementing transitions from the active to the inactive state, and $\gamma$ being the monitoring rate. The coherent dynamics is governed by the Hamiltonian
\begin{equation}
    H = \Omega \sum_{k=1}^{N-1} \left( \sigma_k^x\, n_{k+1} + n_k\, \sigma_{k+1}^x \right) \,,
\end{equation}
with $\sigma^x=\ket{\circ}\!\bra{\bullet}+{\rm h.c.}$ and $n=\ket{\bullet}\!\bra{\bullet}$. 
This operator induces Rabi oscillations at a given site, with rate proportional to $\Omega$, only when at least one of the neighboring sites is in the active state~\cite{buchhold,Marcuzzi2016,roscher}. 
As a consequence of the above dynamical rules, the model features an absorbing state, i.e.~a state which once reached during a trajectory cannot be left. This is the state 
$\ket{0}=\bigotimes_{k=1}^N\ket{\circ}$, which is eventually approached by the system for any finite size $N$. 
In the limit $N\to\infty$ and for sufficiently large ratio $\Omega/\gamma$, the system can instead sustain an active phase characterized by a finite density of active sites~\cite{Non_Eq,buchhold}. The nonequilibrium transition between the absorbing and the active phase in this model has been argued to be a second-order one~\cite{roscher}, as also confirmed by large-scale numerical simulations~\cite{carollo2019}. 

The order parameter of the phase transition is the local density $\langle n_k\rangle$ of active sites. However, the presence of the absorbing state, which is inevitably reached at long times for any \emph{finite system}, poses a key difficulty: the steady-state order parameter becomes uninformative and critical behavior cannot be inferred from stationary properties alone. To overcome this issue, we instead analyze space-time resolved heterodyne trajectories [cf.~Eq.~\eqref{eq:Jhet}], which provide information on the entire evolution of the system. However, since these trajectories have no direct relation to the order parameter and since they are affected by noise, we need an efficient way to extract this information.

\textit{Data generation.---} We generate quantum trajectories by simulating Eqs.~\eqref{eq:SD}--\eqref{eq:Jhet} with matrix product states and a time-evolving block decimation algorithm~\cite{TEBD1,TEBD2}. 
Specifically, for a single trajectory, we represent the wavefunction $\ket{\psi}$ in Eq. \eqref{eq:SD} as a matrix product state. 
In all our simulations, we consider a system of $N=30$ sites, up to time $\gamma t=10$ and use a finite time-step $\gamma {\rm d}t = 0.05$. The ratio $\Omega/\gamma$ is systematically varied and we consider the fully active state $\ket{\psi_0}=\bigotimes_{k=1}^N\ket{\bullet}$ as the initial state of the dynamics. 
In the time-evolving block decimation algorithm, the achievable accuracy is controlled by the bond dimension $\chi$. In our simulations we chose $\chi = 200$, which yielded converged data.

As illustrated in Fig.~\ref{fig:sing_traj}, we compute two types of trajectories. First, we consider the time-dependent expectation value of the local density of active sites, i.e.~$S_k=\bra{\psi} n_k\ket{\psi}$ (trajectories denoted by S in Fig.~\ref{fig:sing_traj}). Subsequently, we focus on the heterodyne trajectories encoding the space-time realizations of the heterodyne current in Eq.~\eqref{eq:Jhet} (trajectories denoted by O in Fig.~\ref{fig:sing_traj}). Both types of trajectories contain space-time dynamical information. The trajectories $S_k(t)$ are, in principle, ideal as they directly display the behavior of the order parameter. However, they cannot be efficiently measured in experiments due to the postselection problem. As we show, 
the heterodyne trajectories $O_k(t)$ are equivalently informative, even if they seem not immediately related to the order parameter and, as shown in Fig.~\ref{fig:sing_traj}, they show no apparent structure throughout the phase diagram.

\textit{Machine-learning methods.---} 
In order to extract information from the noisy heterodyne trajectories $O_k(t)$ [cf.~Fig.~\ref{fig:sing_traj}] and to identify the phase-transition point, we exploit {\it autoencoders}. These are natural tools to learn effective order parameters, as they map the high-dimensional data to a low-dimensional latent space capturing their dominant features~\cite{Deep}. When trained across a phase transition, the latent representation exhibits emergent clustering that separates the two phases. We employ an architecture that takes entire space-time trajectories as input and compresses them to a two-dimensional latent space [see sketch in the left panels of Fig.~\ref{fig:Clustering}]. For the network architecture considered here, a one-dimensional latent representation did not yield a clear correlation between the latent variable and the control parameter $\Omega/\gamma$. While this does not preclude the existence of a suitable one-dimensional representation in general, the emergence of a higher-dimensional effective order parameter in our analysis likely reflects the rich structure of full space-time trajectories compared to steady-state properties. This, in turn, necessitates more flexibility in the learned representation than can be captured by a single number. Details on the architecture and the training of the autoencoder are provided in the End Matter~\cite{SM}.

We train the autoencoder on a total of $1000$ trajectories generated at integer values of the control parameter $\Omega/\gamma\in\{1,2,\dots,10\}$. Once the model has learned to compress the trajectories into a structured, low-dimensional representation, 
we identify clusters in the latent space of the autoencoder [see examples in the central panels of Fig.~\ref{fig:Clustering}] by employing the Gaussian mixture model~\cite{mclachlan2000finite}. The Gaussian mixture model represents complex distributions as a weighted sum of multiple (here two) multivariate Gaussian components. This allows for soft assignments, where points can belong to multiple clusters with varying probabilities~\cite{GMM}. In this way, the network can assign a cluster label, and therefore a phase, to each quantum trajectory it encodes. To assess the consistency of this classification, we sweep $\Omega/\gamma\in[1,10]$ and apply the trained model to $100$ independent trajectories at each value, computing the mean cluster assignment to the active phase. This evaluation includes values of $\Omega/\gamma$ not used during training. As can be seen in the right panel of Fig.~\ref{fig:Clustering}, the assignments are consistent, i.e. they interpolate the results obtained from the training data.

\textit{Classification using a system observable.---}
To test the capability of our model to classify phases, we first consider as input the space-time trajectories $S_k(t)$ of the local order parameter. We independently trained the autoencoder multiple times and here we discuss results from a representative training instance.
After training, the model parameters remain fixed. We then encode individual trajectories and obtain for each of them the predicted probability that it belongs to the active phase. The resulting latent representation of a total of 1000 trajectories exhibits a clear color gradient as shown in Fig.~\ref{fig:Clustering}(a), indicating that the autoencoder can successfully separate trajectories. The following classification yields average cluster assignments that place the critical point within the range $(\Omega/\gamma)^\text{AE}_c\in[5.5,6.5]$.
This interval is just slightly lower than the estimated critical value of $5.9\leq(\Omega/\gamma)_c\leq7$~\cite{carollo2019}. This shows that the autoencoder yields reliable predictions, consistently identifying the transition region. In the vicinity of the critical point, the probability to be associated with the active phase $p_\text{AE}$ is expected to obey a power-law scaling, $p_\text{AE}\propto|\Omega/\gamma-(\Omega/\gamma)^\text{AE}_c|^{\beta^\text{AE}}$, characteristic of a continuous phase transition. Here, $\beta^\text{AE}$ is a critical exponent and a power-law fit to the transition region yields an estimated critical value of $(\Omega/\gamma)^\text{AE}_c=5.8\pm0.3$ and a critical exponent of $\beta^\text{AE}=0.33\pm0.13$~\cite{SM}.

\textit{Classification using the output signal.---}
Finally, we apply the autoencoder and clustering algorithm to heterodyne trajectories.  
To slightly mitigate the noise, we compute the absolute value of the time-averaged current $\overline{O_k(t)}$, where the bar indicates the sliding average over a short time window of length $1/(2\gamma)$. The resulting latent space representation is presented in Fig.~\ref{fig:Clustering}(b). A sharp crossover between the two phases is apparent, placing the critical point in the range $(\Omega/\gamma)^\text{AE}_c\in[5.0,6.5]$. The experimentally accessible quantum trajectories of heterodyne detection therefore allow to pinpoint the transition with an accuracy and precision that is comparable to the experimentally inaccessible quantum trajectories of the local order parameter. A power-law fit to the transition region yields an estimated critical value of $(\Omega/\gamma)^\text{AE}_c=5.4\pm0.1$ and a critical exponent of $\beta^\text{AE}=0.28\pm0.05$~\cite{SM}.

\textit{Conclusions.---}
We have demonstrated that an autoencoder-based clustering algorithm can successfully identify nonequilibrium phase transitions and estimate critical points. We achieved this by providing, as input to the neural network, space-time resolved data emulating a heterodyne continuous-monitoring experiment. Our method therefore solely relies on a readily accessible signal, i.e.~quantum trajectories which are experimentally available in situ and which may not be related to order parameters. 
In the future, it would be interesting to investigate how our autoencoder-based approach can be utilized for analyzing collective behavior in quantum circuits with mid-circuit measurements~\cite{Marcel_ancilla,anand2024,white2026}. In this framework, quantum trajectories associated with heterodyne detection can be emulated by coupling the system with auxiliary two-level systems and measuring their coherences rather than their populations~\cite{gross2018}. Moreover, linking to specific experimental platforms will require the consideration of the impact of non-perfect detection of heterodyne currents on the performance of our machine-learning approach.

% -------------------------------------------------------------------------------------------------------------------------------------------- %

\acknowledgments
\textit{Acknowledgements.---}
We acknowledge funding from the Deutsche Forschungsgemeinschaft (DFG, German Research Foundation) under Germany's Excellence Strategy -- EXC-2111 -- 390814868, through the Research Unit FOR 5413, Grant No. 465199066, and through the Research Unit FOR 5522, Grant No. 499180199. This work was supported by the QuantERA II Programme (project CoQuaDis, DFG Grant No. 532763411) that has received funding from the EU H2020 research and innovation programme under GA No. 101017733. We acknowledge the financial support by the German Federal Ministry of Research, Technology and Space (BMFTR) within the project "Neuronale Quantennetzwerke auf NISQ-Quantencomputern (NeuQuant)" under grant 13N17065. Support was also received through the ERC grant OPEN-2QS (Grant No. 101164443). The authors acknowledge support by the state of Baden-Württemberg through bwHPC and the German Research Foundation (DFG) through grant no INST 40/575-1 FUGG (JUSTUS 2 cluster).
\bibliography{biblio}

% -------------------------------------------------------------------------------------------------------------------------------------------- %

\clearpage
% -------------------------------------------------------------------------------------------------------------------------------------------- %
\section*{End Matter}
\begin{figure}[b]
    \begin{minipage}{\linewidth}
        \centering
        \includegraphics[width=\textwidth]{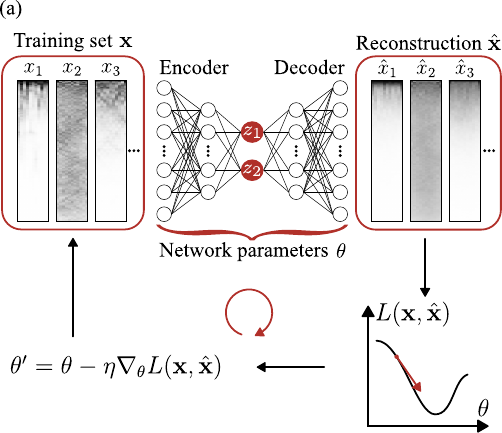}
    \end{minipage}
    \begin{minipage}{0.3\textwidth}
        \flushleft
        \text{(b)}
        
        \centering
        \begin{tabular}{l c}
            \hline\hline
            Parameter & Value \\
            \hline
            Learning rate $\eta$ & $0.001$ \\
            Number of epochs & $20$ \\
            Training samples & $1000$ \\
            Batch size & $10$ \\
            Optimizer & Adam \\
            Activation function & ReLU \\
            Loss function $L$ & MSE \\
            \hline\hline
        \end{tabular}
        \vspace{3.5mm}
    \end{minipage}
    \caption{\textbf{Neural network architecture and training.} (a) Architecture of the autoencoder used to compress the full space-time trajectories into a two-dimensional latent representation $z_1,z_2$. The network consists of layers with dimensions $6000 \times 1000 \times 2 \times 1000 \times 6000$. For the time-averaged heterodyne current the model size is reduced to $5730 \times 1000 \times 2 \times 1000 \times 5730$ nodes. During training, trajectories $\mathbf{x}=(x_1,x_2,\dots)$ are processed by the autoencoder, and the network parameters $\theta$ are updated at each iteration to minimize the reconstruction loss $L(\mathbf{x},\hat{\mathbf{x}})$ between the input $\mathbf{x}$ and its reconstruction $\hat{\mathbf{x}}$. The learning rate $\eta$ controls the step size of parameter updates $\theta\to\theta^\prime$ during optimization. (b) Summary of the training parameters.}
    \label{fig:Sketch_Architecture}
\end{figure}

\textit{Autoencoder: Architecture and training.---}
Autoencoders are a type of neural network designed for unsupervised learning. They consist of two main parts: The \textit{encoder}, which compresses the input data into a lower-dimensional latent space, and the \textit{decoder}, which tries to reconstruct the original data from the compressed representation. The goal is for the autoencoder to learn compact representations of the data, making it useful for tasks like dimensional reduction, denoising, and anomaly detection~\cite{Deep}. Training an autoencoder amounts to optimizing the network parameters so as to minimize the reconstruction loss $L(\mathbf{x},\hat{\mathbf{x}})$ over a training set $\mathbf{x}=(x_1,x_2,\dots)$, where $\hat{x}_i$ denotes the reconstruction of $x_i$. To achieve this, the backpropagation algorithm~\cite{Backprop} is employed to calculate the gradient of the loss function with respect to the network's parameters $\mathbf{\theta}$. The latter are then adjusted in the opposite direction of the gradient, $\mathbf{\theta}^\prime=\mathbf{\theta}-\eta\nabla_\mathbf{\theta} L(\mathbf{x},\hat{\mathbf{x}})$, where $\eta$ denotes the so-called learning rate. Over multiple iterations (epochs), this process allows the network to converge towards a local minimum of the loss function, ideally leading to improved performance on the task. The employed architecture and the training routine are illustrated in Fig.~\ref{fig:Sketch_Architecture}(a).
Our models were trained for 20 epochs using 1000 different trajectories, with a learning rate of $\eta=0.001$. This learning rate was determined as optimal after conducting a grid search. The number of epochs chosen was sufficient, as the loss function converged quickly. Although various architectures were explored, the simpler design depicted in Fig.~\ref{fig:Sketch_Architecture}(a) delivered the best performance. All training parameters are summarized in Fig.~\ref{fig:Sketch_Architecture}(b). 
Extending the autoencoder using recurrent layers~\cite{recurrent} or adopting transformer-based models~\cite{attention} may allow for improvements on the accuracy and on the interpretability of the inferred transition behavior.

\begin{figure}
    \centering
    \includegraphics[width=0.85\linewidth]{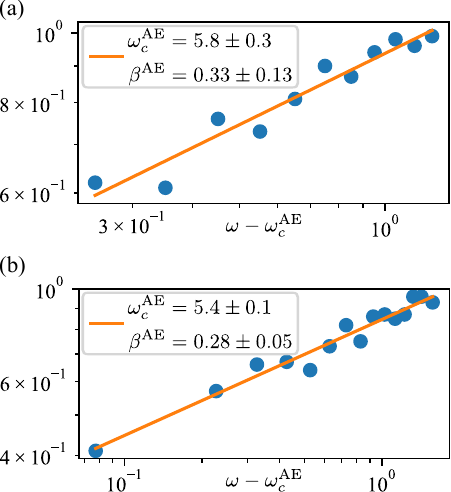}
    \caption{\textbf{Critical scaling at the quantum contact process transition.} (a) Probability of being associated with the active phase, extracted from trajectories of the local density of active sites $S_k(t)$ [see Fig.~\ref{fig:Clustering}(a)]. A power-law behavior is fitted to the data in the interval $\omega\in[6.0,7.0]$. (b) Probability of being associated with the active phase, extracted from trajectories of the absolute value of the time-averaged heterodyne current $|\overline{O_k(t)}|$ [see Fig.~\ref{fig:Clustering}(b)]. A power-law fit is performed over the interval $\omega\in[5.5,7.0]$. In both cases, the critical exponent $\beta^\text{AE}$ and the critical point $\omega^\text{AE}_c$ are treated as free parameters.}
    \label{fig:beta_fit}
\end{figure}

\textit{Estimation of the critical point and critical exponent.---}
Let $\omega=\Omega/\gamma$ be the dimensionless control parameter of the quantum contact process. Close to the critical point $\omega_c$, one may expect the probability to be associated with the active phase $p_\text{AE}$ to obey a power-law scaling, $p_\text{AE}\propto|\omega-\omega^\text{AE}_c|^{\beta^\text{AE}}$, where $\beta^\text{AE}$ is a critical exponent. To explore this scaling behavior, we investigate a range of $\omega$ close to the crossover region and perform a power-law fit with $\beta^\text{AE}$ and $\omega^\text{AE}_c$ treated as free parameters. The results are shown in Fig.~\ref{fig:beta_fit}. We observe that the position of the critical point is in good agreement with previous studies of the quantum contact process~\cite{carollo2019,Kahng2021}. Moreover, the critical exponent that characterizes the scaling behavior of the probability of being in the active phase appears to be close to the static scaling exponent of the order parameter that was also previously found for the quantum contact process. We consistently observe  this across different training instances in both the local density of active sites $S_k(t)$ and the heterodyne current $O_k(t)$. This robustness suggests that the autoencoder captures features governed by the underlying critical behavior.

\end{document}